\documentclass{cpbtex}
\usepackage{array}
\usepackage{float}
\usepackage{booktabs}
\usepackage{textcomp}
\usepackage{multirow}
\usepackage{amstext}
\usepackage{graphicx}
\usepackage{color}
\usepackage{ulem}

\begin{document}

\title{Dynamical signatures of the one-dimensional deconfined quantum critical point\thanks{Project supported by the National Science Foundation of China (Grant No. 12174441) and the Fundamental Research Funds for the
Central Universities and the Research Funds of Remnin University of China (Grant No. 18XNLG24).}}


\author{Ning Xi$^{1}$, \ and \ Rong Yu$^ {1}$\thanks{Corresponding author. E-mail:rong.yu@ruc.edu.cn}\\
$^{1}$Department of Physics and Beijing Key Laboratory of Opto-electronic Functional Materials\\
 and Micro-nano Devices, Renmin University of
China, Beijing 100872, China}  

\date{\today}
\maketitle

\begin{abstract}
We study the critical scaling and dynamical signatures of fractionalized excitations at two different deconfined quantum critical points (DQCPs) in an $S = 1/2$ spin chain by using the time evolution of infinite matrix product states. The scaling of the correlation functions and the dispersion of the conserved current correlations explicitly show the emergence of enhanced continuous symmetries at these DQCPs. The dynamical structure factors in several different channels reveal the development of deconfined fractionalized excitations at the DQCPs. Furthermore, we find an effective spin-charge separation at the DQCP between the ferromagnetic (FM) and valence bond solid (VBS) phases, and identify two continua associated to different types of fractionalized excitations at the DQCP between the $X$-direction and $Z$-direction FM phases. Our findings not only provide direct evidence for the DQCP in one dimension but also shed light on exploring the DQCP in higher dimension.
\end{abstract}



\section{Introduction}

Significant quantum fluctuations in low-dimensional spin frustrated systems
can give rise to 
novel quantum phases and 
exotic quantum phase transitions~\cite{Chaikin_1995, Giamarchi_book:2004, diep_2004, lacroix_mendels_mila_2013, Sachdev_book:2011}. One substantial difference from the classical systems is that in these quantum spin systems, some phases can not be characterized by conventional order parameters with broken symmetry. Correspondingly, quantum phase transitions may beyond the standard Landau-Ginzburg-Wilson (LGW) paradigm. Surprisingly, even when the ordered phases can be described by conventional order parameters with spontaneous symmetry breaking (SSB), the quantum phase transition between them may still be exotic and beyond the LGW paradigm. One of the most attractive examples is the DQCP between the N\'{e}el antiferromagnetic (AFM) and VBS states in two dimension, proposed by Senthil et al~\cite{DQCP2004, Senthil2004}. Landau theory dictates a first-order VBS-AFM transition because the order parameters of the two phases break different symmetries. But the theory of DQCP predicts a continuous phase transition between these two phases. At the DQCP, deconfined fractionalized spin excitations emerge and enhanced symmetry allows continuous rotation between the order parameters of the two ordered phases.

The DQCP in two-dimensional (2D) systems has been extensively studied theoretically~\cite{Sandvik2007, Chen_PRL:2013, Nahum_PRL:2015, Shao_Science:2016, Meng2017, Meng2018, Meng2019, Lee2019, Xi2021firstorder, Bowen2020, Guangyu2020}. Although enormous computational cost are needed to clarify the unusual critical properties and deconfined fractionalized excitations emergent at the DQCP, numerical evidences are cumulated. Currently, most studies are based on several sophisticatedly designed models.
To realize a DQCP in realistic 2D spin models or even in experimental quantum magnets~\cite{Cui2021} 
is, however, still a challenging task.
Recently, some works studied possible DQCP 
in one-dimensional (1D) spin systems and obtained 
some interesting results~\cite{Huang2019, Huang2020, Sandvik2018, Toshiya2019, MotrunichNUM2019, MotrunichTHE2019, Furusaki2012,Sheng2021,arXiv02821,PhysRevB.103.155143,Affleck1987,Luo2019,Sun2019}.
An advantage of the 1D quantum spin systems over their 2D counterparts is that many powerful and well controlled
analytical and numerical techniques, such as Bethe-ansatz, bosonization, and density matrix renormalization group (DMRG) \cite{fradkin_2013, Bosonization, DMRG} can be applied, and hence the results are more convincing. Meanwhile, enhanced quantum fluctuations in 1D employ strong constraint to ground-state properties, as involved in the Lieb-Schultz-Mattis (LSM) theorem~\cite{LIEB1961407,LSM2015,LSM2017}. It is recently noticed that this may help stabilizing a DQCP in 1D~\cite{MotrunichTHE2019}.

For an $S=1/2$ chain with $\text{SO(3)}$ spin rotational and lattice translational symmetries, the LSM theorem~\cite{LIEB1961407} implies that the ground state either breaks the translational symmetry to form a VBS or keeps plainly gapless. When generalized to systems with discrete symmetries, this theorem dictates that the ground state can not be symmetric and plainly gapped. Let us consider a system with two distinct discrete symmetries and assume a single tuning parameter can drive the system from one SSB phase to the other. At the transition point the symmetries restore and are often enhanced to allow continuous rotation between the two order parameters. According to the LSM theorem, the ground state at this point must be gapless, and in 1D, this implies the transition point is a DQCP. In many systems, the microscopic Hamiltonian contains such a symmetry-enhanced point by properly tuning the model parameter. For example, the XY model consists of an isotropic point with continuous spin $\text{U(1)}$ symmetry by tuning the spin anisotropy. Constrained by the LSM theorem, in 1D models, these symmetry-enhanced points exhibit properties that are eligible for a DQCP. This is a special case in 1D, as discussed later in this paper. In another class of systems, the enhanced symmetry can not be explicitly read off from the microscopic Hamiltonian, but emerges as both SSB order parameters are simultaneously suppressed by fluctuations. At the same time deconfined fractionalized excitations emerge, making this symmetry-enhanced point a DQCP. This case is more generally discussed in both 1D and 2D.

In this work, we investigate the dynamical signatures of deconfined fractionalized excitations and enhanced
continuous symmetry of two types of DQCP in 1D by studying an
$S=1/2$ spin chain with FM nearest-neighbor and AFM next nearest-neighbor interactions. We calculate the static and dynamic spin correlations by using the time evolution of infinite matrix product states (MPS). We show the existence
of emergent continuous symmetries and characteristic of fractionalized excitations
at two different DQCPs. At the DQCP between the VBS and FM phases, our calculation of dynamic structure factors suggests there is an effective spin-charge separation. While at the DQCP between the $X$-direction
and $Z$-direction FM phases, we observe two different types of continua associated to different deconfined fractionalized excitations.
Our results show direct evidence for the DQCP in 1D and provide valuable information in understanding the nature of the DQCP in both 1D and 2D.

\section{Model and Methods}

\begin{figure}[!t]
 \centering
\includegraphics[width=15cm]{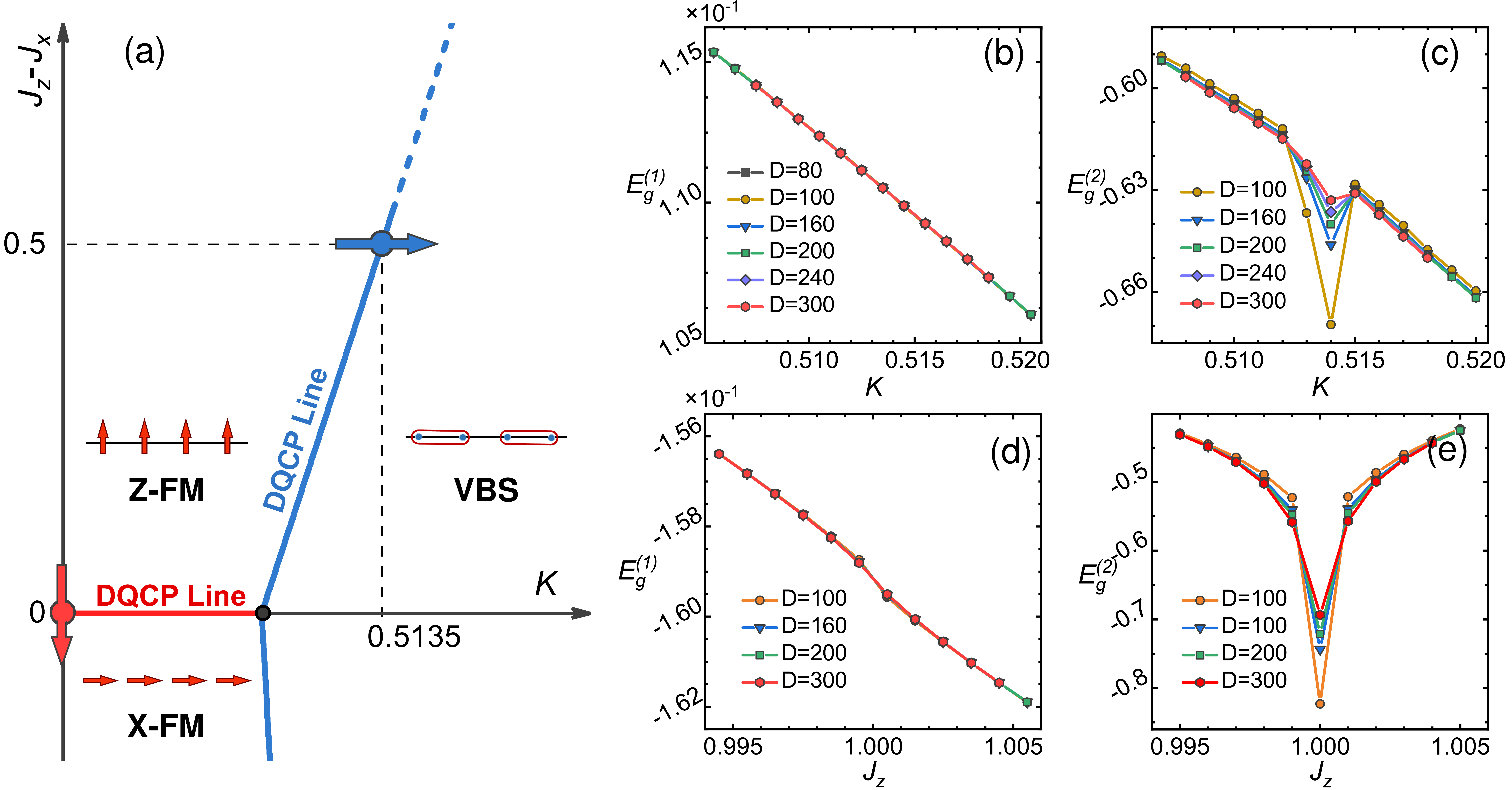}
\caption{(a) A schematic phase diagram for the model in Eq.~\eqref{eq:spinmodel}. The isotropic line
with quasi-long-range(QLR) order (red line) separates the $X$-direction
and $Z$-direction ferromagnetic (FM) phases, i.e., a DQCP line.
The DQCP line between the FM and VBS phases (blue solid line) with
emergent $\text{O(2)}$\texttimes $\text{O(2)}$ symmetry has been studied in Ref. \cite{Huang2019}.
A more complete phase diagram is mentioned in Ref. \cite{MotrunichNUM2019}. Here, we
just focus on these two deconfined transitions (go through red line
or blue line) without examining the properties over the entire phase
diagram. (b) The first derivative $E_{g}^{(1)}$ of the ground-state
energy with $K$ along the blue arrow in panel (a). (c) The second
derivative $E_{g}^{(2)}$ of the ground-state energy with $K$ along
the blue arrow in panel (a). (d) and (e) Same to (b) and (c) but along
the red arrow.}
\label{fig:1}
\end{figure}

To study the properties of DQCP in 1D, we consider an $S=1/2$ spin chain with the following Hamiltonian:
\begin{equation}
H=\sum_{i}\left(-J_{x}S_{i}^{x}S_{i+1}^{x}-J_{z}S_{i}^{z}S_{i+1}^{z}\right) +K\left(S_{i}^{x}S_{i+2}^{x}+S_{i}^{z}S_{i+2}^{z}\right).\label{eq:spinmodel}
\end{equation}
Here $J_x$ and $J_z$ are nearest-neighbor FM exchange interactions and $K$ refers to the next nearest-neighbor AFM exchange coupling. This model has the $\mathbb{Z}_2^{x}\times \mathbb{Z}_2^{z}$ spin rotational symmetry, as well as the time reversal and lattice translational symmetries. It supports an $X$-direction FM, a $Z$-direction FM, and a VBS in the ground-state phase diagram as depicted in Fig.~\ref{fig:1}(a). In this work, we set $J_{x}=1$ as the energy unit and study two quantum phase transitions: one along the blue arrow (by fixing $J_{z}=1.5$ and increasing $K$) and the other along the red arrow (by fixing $K=0$ and
decreasing $J_{x}$), as shown in Fig.~\ref{fig:1}(a), respectively.

This model has been analyzed by using several different approaches~\cite{MotrunichNUM2019, MotrunichTHE2019, Furusaki2012, Huang2019}. Ref.~\cite{Furusaki2012} has numerically confirmed that the ground states include a Luttinger liquid, a VBS phase, and a vector chiral phase sequently by increasing $K$ along the isotropic line ($J_x=J_z$). Various field theory descriptions of the DQCPs between the FM and VBS phases have been discussed in detail in Ref.~\cite{MotrunichTHE2019}. Another interesting analogy is mentioned in Refs.~\cite{MotrunichTHE2019,Furusaki2012} that the VBS phase can be viewed as a symmetry-protected topological (SPT) phase by considering unit cells of two lattice sites. However, this SPT-like phase is obviously topologically trivial when examining the edge mode, given that the VBS is a SSB phase. But for the case of an infinite chain, the edge is unimportant, and therefore this analogy can make senses.

 An anisotropic $\text{O(4)}$ non-linear sigma model (NLSM) with the Wess-Zumino-Witten term has been proposed to describe the scaling
behaviors and emergent symmetry of the DQCP between the FM and VBS phases~\cite{Huang2019}.
Here, to gain further understanding on the symmetry and fractionalized excitations of the two DQCPs, we apply the celebrated Jordan-Wigner (JW) transformation to this spin model and map it to interacting spinless fermions. The Hamiltonian in the fermion representation 
then reads:
\begin{equation}
H=H_{\mathrm{0}}+H_{\mathrm{int}},
\end{equation}
where
\begin{equation}
H_{\mathrm{0}}=-\frac{J_{z}+J_{x}}{4}\sum_{i}\hat{T}_{i} -\frac{J_{z}-J_{x}}{4}\sum_{i}\left(f_{i}^{\dagger}f_{i+1}^{\dagger}+f_{i+1}f_{i}\right),
\end{equation}
and
\begin{equation}\label{eq:Hint}
H_{\mathrm{int}}=-\frac{K}{2}\sum_{i}\left(\hat{T}_{i}-\hat{T}_{i+1}\right)^{2} +K\sum_{i}\left(\hat{N}_{i}-\hat{N}_{i+1}\right)^{2}.
\end{equation}
Here $f_{i}$ is the annihilation operator for fermions on site $i$,
$\hat{T}_{i}$ and $\hat{N}_{i}$ refer to the nearest-neighbor hopping
and on-site particle number operators, respectively, defined as
\begin{equation}
\hat{T}_{i}=f_{i}^{\dagger}f_{i+1}+f_{i+1}^{\dagger}f_{i}=2\left(S_{i}^{x}S_{i+1}^{x}+S_{i}^{z}S_{i+1}^{z}\right),
\end{equation}
and
\begin{equation}
\hat{N}_{i}=f_{i}^{\dagger}f_{i}=\left(\frac{1}{2}+S_{i}^{y}\right).
\end{equation}
$H_{\mathrm{0}}$ corresponds to the nearest-neighbor term of Eq.
\eqref{eq:spinmodel}. The interaction term $H_{int}$ is transformed from the next nearest-neighbor
term of the spin model. Note that $\hat{T}_{i}$ is defined ``on-bond'' and
$\hat{N}_{i}$ is defined ``on-site''.

In this work, we adopt the infinite time-evolving block decimation
(iTEBD)\cite{Vidal,SCHOLLWOCK2011} method to study the ground-state
properties and zero-temperature space-time correlations of the spin
model defined in Eq.~\eqref{eq:spinmodel}. We focus on the critical behaviors near the two transitions shown in Fig.~\ref{fig:1}(a). The matrix product representation
is known to be accurate and efficient for gapped systems when the truncation dimension
(Schmidt rank) $D$ is large enough. In the situation where the system is gapless, the matrix
product representation can still be applied and it provides a finite-gap (or finite-entanglement)
variational state of a gapless system. With various well developed finite correlation length or finite $D$ scaling techniques, MPS representation is capable of exploring the critical properties
trustingly\cite{scaling2008,scaling2009,scaling2012}.

In the iTEBD method, one usually starts from a random initialization. Deep inside an ordered phase, the ground state is not sensitive to the way of initialization. The local order parameters and ground-state energy
can be determined sufficiently accurate at a finite $D$. However, when the system
is close to a critical point, finite-$D$ effects may appear. There
are usually two issues. The first issue is related to the emergent symmetry at the transition.
Near the transition point, the emergent symmetry will give rise to many nearly degenerate ground state configurations, which are local minima in the variational approach. To avoid being trapped in a local minimum, our optimization procedure starts with many different initializations and the state with the lowest energy in the optimization is selected as the ground state. As a second issue, the gap obtained in a finite-$D$ MPS is always finite. To solve this problem, we apply the well developed
finite entanglement scaling technique, which allows to accurately locate the critical point from extrapolation.

For a state with a finite
$D$, a finite gap presents because of the finite correlation
length $\xi(D)$. To eliminate the effects of the finite $D$ gap to the critical exponents, we adopt a self-consistent
method to determine the critical exponents. Applying the finite correlation length scaling approach, the critical exponents and the scaling function can be obtained from data collapse for different $D$'s. We then use the fitted results to do a power-law fitting for the data with the largest correlation length, from which the critical exponents are reestimated. This procedure then repeats until the evaluated exponents are converged.
Details of this self-consistent scaling method is elaborated in Appendix B.

The dynamical structure factors are calculated from the Fourier transforms
of the space-time correlations. The calculation of space-time correlations
is based on the real-time evolution of the ground-state MPS\cite{White2008}.
The computational cost of calculating the space-time correlations is much higher than
optimizing a ground state. To keep the balance between truncation
error and energy resolution, in this work, we adopt a fourth-order
Trotter expansion of the real-time evolution operator which leads to the energy
resolution less than $10^{-3}J$.

\section{Ground-state phase diagram and symmetries at the DQCPs}

The schematic phase diagram of the model defined in Eq.~\eqref{eq:spinmodel} is presented in Fig. \ref{fig:1}(a). For $J_z>J_x$ and small $K$ the ground state is a FM with spin ordered along the $S^z$ direction (denoted as the $Z$-direction FM). Since the model is symmetric under the $x\leftrightarrow z$ interchange, the ground state is an $X$-direction FM for $J_z<J_x$ and small $K$. For sufficiently large $K$, the ground state is a VBS that breaks the translational symmetry. The $X$-direction and $Z$-direction FM states
meet at the isotropic line $J_z=J_x$ where the spin rotational symmetry of the Hamiltonian is enhanced from $\mathbb{Z}_2^{x}\times \mathbb{Z}_2^{z}$ to $\text{U(1)}$. Constrained by the LSM theorem, along this line, the ground state preserves the $\text{U(1)}$ symmetry and keeps to be gapless. Across this line from the $X$-direction FM to the $Z$-direction FM by tuning the parameter $J_z-J_x$ in the model, the system undergoes a continuous transition. At the transition point, the low-energy excitations are described by JW fermions. The JW fermions are deconfined with the enhanced $\text{U(1)}$
symmetry, and can be viewed as the fractionalized excitations of domain walls (DWs), just like the marons or instantons for vortex in 2D. Therefore, the phase boundary between the $X$- and $Z$-direction FM phases (red line in Fig. \ref{fig:1}(a)) can be regarded as a line of DQCPs.

The transition
between the $Z$-direction FM and VBS phases is driven by the next nearest-neighbor coupling $K$. It has been proposed that DQCPs with emergent $\rm \text{O(2)}\times \rm \text{O(2)}$ symmetry exist along the FM and VBS phase boundary (the blue line in Fig. \ref{fig:1}(a)) ~\cite{MotrunichTHE2019, MotrunichNUM2019, Huang2019}. To understand the origin of the enhanced continuous symmetry, we examine the system in the JW fermion representation. The transition is controlled by $H_{int}$ in Eq.~\eqref{eq:Hint}, which contains an ``on-bond'' term (including $\hat{T}_{i}$) and an ``on-site'' term (including $\hat{N}_{i}$). Relevance of the ``on-bond'' term will cause breaking the translational symmetry on bonds and preserving all on-site symmetries, and therefore form a VBS. By
contrast, the relevance of the ``on-site'' term will cause breaking the
translational symmetry on sites and preserving all on-bond symmetries. This would result in
a $Y$-direction AFM phase. For sufficiently large $K$ in this model, 
the ground state prefers to a VBS instead of a $Y$-AFM state. In the continuous limit,
these two terms of $H_{int}$ 
correspond to the same order of interaction but pinning down different ordered states. There are also two sets of rotational symmetries originated from the ``on-bond" and ``on-site" symmetries, respectively. Interestingly, with proper tuning, the ``on-bond" and ``on-site" terms can be simultaneously marginal, which
cause the enhanced ${\rm \text{O}}(2)\times{\rm \text{O}}(2)$ symmetry.
Note that these emergent continuous symmetries at the transition is not possessed by the microscopic Hamiltonian of Eq.~\eqref{eq:spinmodel}. This is a key difference from the transition across the boundary of the two FM phases. However, due to the LSM theorem in 1D, they share many similar features, and are both eligible for DQCPs.

\section{Critical properties at DQCPs}

\begin{figure}[h]
 \centering
\includegraphics[width=15cm]{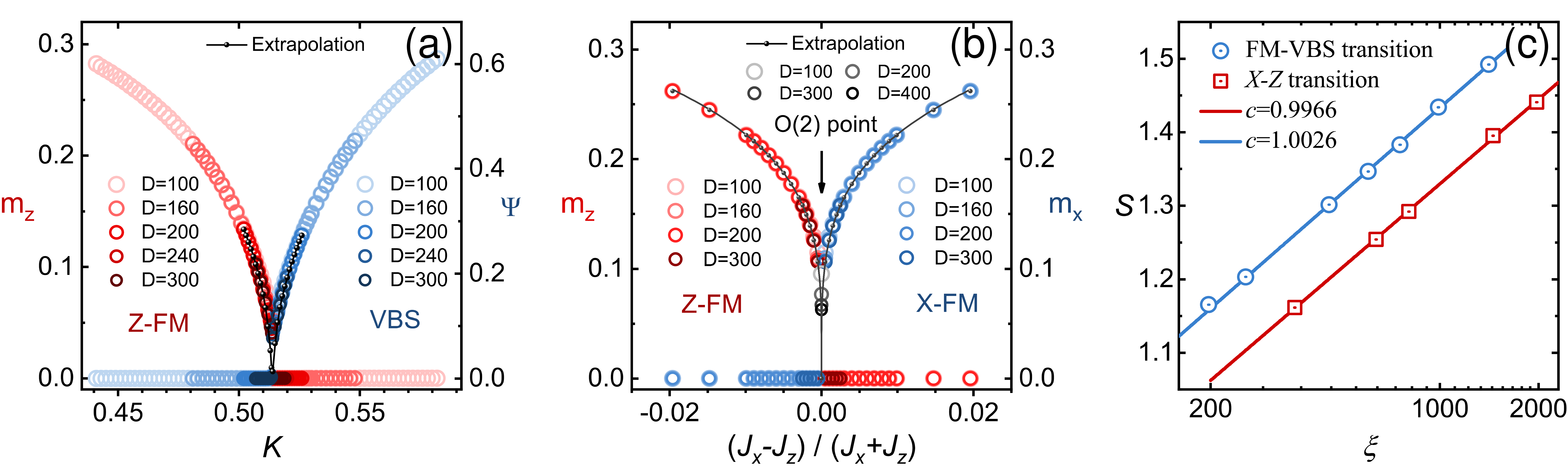}

\caption{(a) $Z$-FM order parameter $m_{z}$ and VBS order parameter $\Psi$
versus $K$ at $J_{z}-J_{x}=0.5$. (b) $Z$-FM order parameter $m_{z}$
and $X$-FM order parameter $m_{x}$ versus $(J_{x}-J_{z})/(J_{x}+J_{z})$
at $K=0$. (c) Scaling of the entanglement entropy $S$ with effective
correlation length $\xi$. The central charge is extracted from the slope of a linear fit. \label{fig:2}}
\end{figure}

We first examine the critical properties of the $Z$-direction FM to VBS transition. The ground-state
energy $E_{g}$, its first derivative $E_{g}^{(1)}=dE_{g}/dK$,
and the second derivative $E_{g}^{(2)}=dE_{g}^{(1)}/dK$ have been calculated. As demonstrated
in Fig. \ref{fig:1} (b) and (c), $E_{g}^{(1)}$ varies continuously
across the transition, and $E_{g}^{(2)}$ develops a singularity at
the transition point($K\thickapprox0.5135$), featuring a continuous
transition. We further calculate the order parameters $m_{z}=\sum_{i}\left\langle S_{i}^{z}\right\rangle /N$
and $\Psi=\sum_{i}\left\langle \mid\Psi_{i}-\Psi_{i+1}\mid\right\rangle /N$ of the Z-FM
and VBS phases, respectively, to verify the continuous nature of the
transition. A more concrete definition of the order parameters is
listed in Tab. \ref{tab:1-2}. As shown in Fig. \ref{fig:2}(a), $m_{z}$
and $\Psi$ both exhibit near-continuous transitions at the same point,
consistent with the results of ground-state energy, giving robust
evidence of a DQCP. It is worth noting that at any finite-$D$, both order parameters take small finite values at the transition point. This finite-$D$ effect is inherent in the MPS
approximation of the long-range entangled gapless state, as
mentioned in the previous section. To obtain the behavior of order parameters in the thermodynamic limit, we extrapolate their values in the large-$D$ limit, as also shown in Fig. \ref{fig:2} (a). Indeed, both order parameters are continuously suppressed to zero at the same $K$ value.

Fig. \ref{fig:1}(d) and (e) show the derivatives of ground-state energy across the $X$-FM to $Z$-FM transition. The results are similar to those in (b) and (c). Meanwhile, the order parameters of the two SSB
phases, $m_{z}$ and $m_{x}$, both
drop down to zero at the transition point simultaneously, as shown in Fig. \ref{fig:2}(b).
These results evidence a continuous transition, and imply that the symmetry is enhanced to $\text{U(1)}$ at the transition point, which is consistent with the LSM theorem. Besides the similar behaviors in the order parameters, we find the central charges of the two transitions are both close to $1$ (see Fig.~\ref{fig:2}(c)), which belong to the
free boson universality class\cite{Latorre_2009,entanglefootnote}. This does not mean the low-energy excitations are bosonic, but implies that they arise from fractionalizing the bosonic modes, which is a feature of DQCP.

To further show the two transitions are eligible for DQCPs, we further calculate the correlation
functions to verify the existence of emergent continuous symmetry
and global topological conservation. The gauge constraint at the DQCPs
guarantees the conservation of the Noether current, $\partial_{\mu}\mathcal{J^{\mu}}=0$, which implies
that the anomalous exponents of the conserved currents must be pinned
to $\eta=2$, and the current-current correlation in energy-momentum
space must be gapless and linear.

\begin{table}[H]
 \centering
\caption{Identification of the vector components and conserved currents for different
emergent $\rm \text{O(2)}$ symmetries.}
\label{tab:1-2}
\begin{tabular}{>{\raggedright}m{1.4cm}>{\raggedright}m{1.4cm}>{\centering}m{5.5cm}>{\raggedright}m{3.5cm}}
\toprule
\centering{}{\footnotesize{}Transition} & {\footnotesize{}Channel} & {\footnotesize{}Vector components} & \centering{}{\footnotesize{}Conserved current}\tabularnewline
\toprule
\multirow{4}{1.5cm}{\centering{}{\footnotesize{}FM-VBS}} & \multirow{2}{1.5cm}{{\footnotesize{}${\rm \text{O(2)}}_{c}$}} & {\footnotesize{}$S_{i}^{z,u}=S_{i}^{z}+S_{i+1}^{z}$} & \multirow{2}{4cm}{\centering{}{\footnotesize{}$\mathcal{J}_{i}^{c}=S_{i}^{x}S_{i+1}^{y}+S_{i}^{y}S_{i+1}^{x}$}}\tabularnewline
 &  & {\footnotesize{}$\Psi_{i}=S_{i}^{x}S_{i+1}^{x}+S_{i}^{z}S_{i+1}^{z}-S_{i}^{y}S_{i+1}^{y}$} & \tabularnewline\addlinespace[0.2cm]
\cmidrule{2-4} \cmidrule{3-4} \cmidrule{4-4}
 & \multirow{2}{1.5cm}{{\footnotesize{}${\rm \text{O(2)}}_{s}$}} & {\footnotesize{}$S_{i}^{x,u}=S_{i}^{x}+S_{i+1}^{x}$} & \multirow{2}{4cm}{\centering{}{\footnotesize{}$\mathcal{J}_{i}^{s}=S_{i}^{z}-S_{i+1}^{z}$}}\tabularnewline
 &  & {\footnotesize{}$S_{i}^{y,s}=S_{i}^{y}-S_{i+1}^{y}$} & \tabularnewline\addlinespace[0.2cm]
\toprule
\multirow{5}{1.5cm}{\centering{}{\footnotesize{}X-Z}} & \multirow{3}{1.5cm}{{\footnotesize{}${\rm \text{SU(2)}}_{c}$}} & {\footnotesize{}$S_{i}^{y,u}=S_{i}^{y}+S_{i+1}^{y}$} & \multirow{3}{4cm}{\centering{}{\footnotesize{}$\Phi_{i}$, $\Gamma_{i}$, and $S_{i}^{y,u}$}}\tabularnewline
\addlinespace[0.2cm]
 &  & {\footnotesize{}$\Gamma_{i}=S_{i}^{z}S_{i+1}^{x}+S_{i}^{x}S_{i+1}^{z}$} & \tabularnewline\addlinespace[0.2cm]
 &  & {\footnotesize{}$\Phi_{i}=S_{i}^{x}S_{i+1}^{x}-S_{i}^{z}S_{i+1}^{z}$} & \tabularnewline\addlinespace[0.2cm]
\cmidrule{2-4} \cmidrule{3-4} \cmidrule{4-4}
 & \multirow{2}{1.5cm}{{\footnotesize{}${\rm \text{O(2)}}_{s}$}} & {\footnotesize{}$S_{i}^{z}$} & \multirow{2}{4cm}{\centering{}{\footnotesize{}$\mathcal{J}_{i}^{s}=S_{i}^{y}$}}\tabularnewline
 &  & {\footnotesize{}$S_{i}^{x}$} & \tabularnewline
\bottomrule
\end{tabular}
\end{table}

\begin{figure}[h]
 \centering
\includegraphics[width=7.5cm]{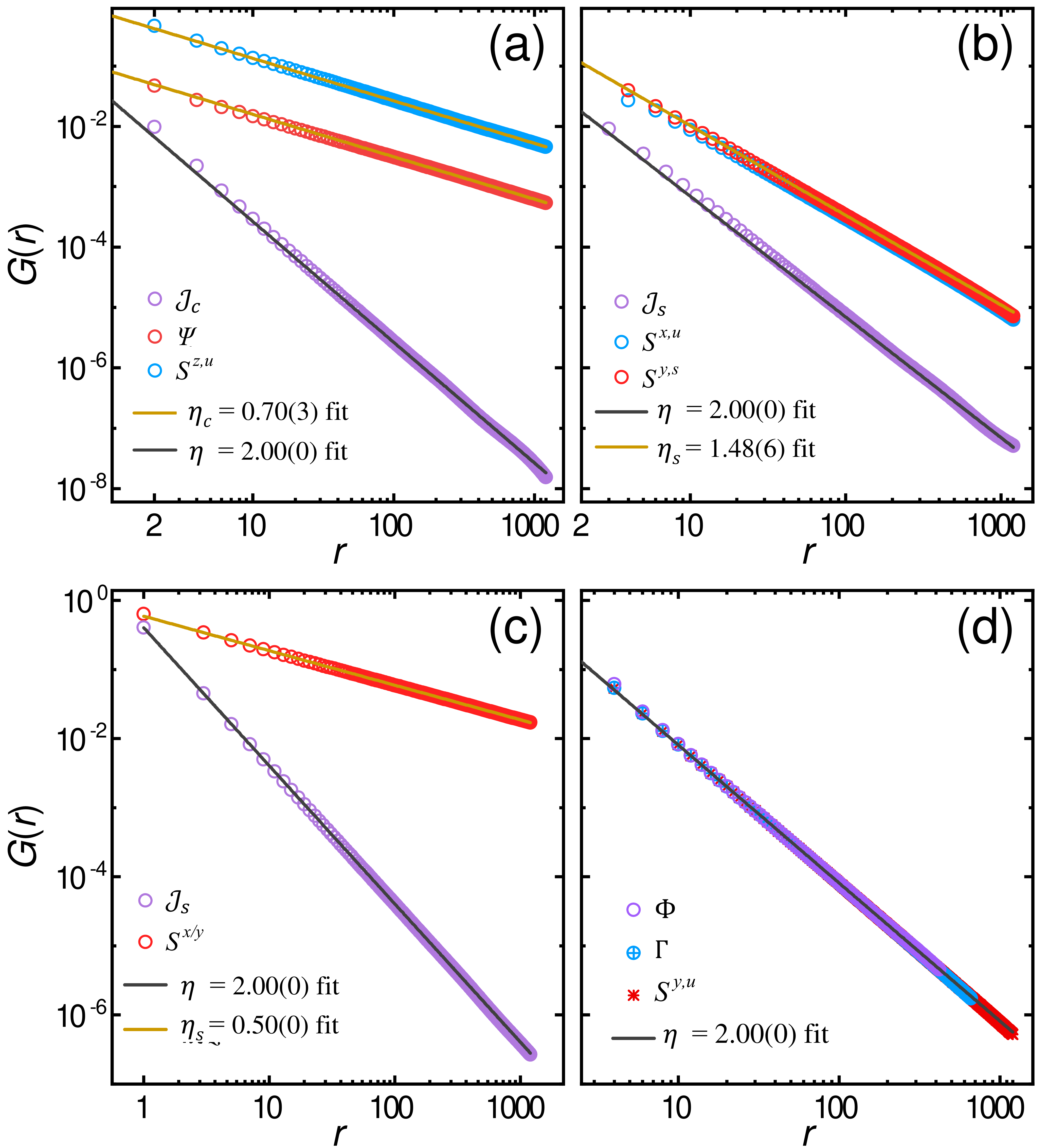}

\caption{Correlation functions of the emergent conserved currents and vector
components for (a) ${\rm \text{O(2)}}_{c}$ of the FM-VBS transition, (b) ${\rm \text{O(2)}}_{s}$
of the FM-VBS transition, (c) ${\rm \text{O(2)}}_{s}$ of the $X$-$Z$ FM transition,
and (d) ${\rm \text{SU(2)}}_{c}$ of the $X$-$Z$ FM transition.\label{fig:3}}
\end{figure}

The vector components and conserved currents of emergent symmetries are listed in Tab. \ref{tab:1-2}. For the FM-VBS DQCP, it has been proposed that there are two emergent $\rm \text{O(2)}$ symmetries~\cite{Huang2019}. In the spinon
language, the continuous rotation of uniform $S^{x}$ and staggered
$S^{y}$ is the gauge symmetry on site with spin flip and the rotation
of $S^{z}$ and $\Psi$ is the gauge symmetry on bond without spin
flip. Thus, we can regard the rotation of uniform $S^{x}$ and staggered
$S^{y}$ as an effective ``spin'' sector, labeled as $\text{O(2)}_{s}$.
On the other hand, the rotation of $S^{z}$ and $\Psi$ form
an effective ``charge'' sector, labeled as $\text{O(2)}_{c}$.
As shown in Fig \ref{fig:3}(a) and (b), although these
two different $\rm \text{O(2)}$ symmetries keep different anomalous exponents,
their conserved currents both follow a scaling law of $\mathcal{J}\sim r^{-2}$
and in each sector (charge or spin) the two vector components behave with the same
scaling dimension. Besides, an additional result $\eta_{c}*\eta_{s}\approx1$
validates the prediction of field theory in Ref \cite{Huang2019}.
For the transition between the two FM states at the isotropic point with $J_{x}=J_{z}$, there is an enhanced $\text{U(1)}\sim \text{O(2)}$ symmetry. As demonstrated in Fig. \ref{fig:3} (c),
the scaling law of $S_{y}\sim r^{-2}$ reflects the global conservation
of the JW fermion number. It can also be regarded as the conserved current
of the $X$-$Z$ rotation, labeled as $\text{O(2)}_{s}$. To further illustrate
the notion of deconfinement, we show that another $\text{SU(2)}_{c}$ symmetry listed
in Tab. \ref{tab:1-2} is also emergent at the DQCP of $X$-$Z$ transition.
As demonstrated in Fig. \ref{fig:3} (d), the anomalous exponents
of these three operators are all pinned at an integer of 2, which
indicates an emergent $\text{SU(2)}$ symmetry.

\section{Dynamical Signatures of DQCPs}\label{Sec:5}

Results in previous sections only provide evidences for continuous FM-FM and FM-VBS transitions, based on which the DQCPs are implied. To directly show the existence of deconfined fractionalized excitations at the DQCPs, we calculate the space-time correlations of conserved currents and spin components
by real-time evolution of the ground-state MPS. After Fourier transforms
of the space-time correlations, we can get a dynamical spectrum in
the energy-momentum space.

\begin{figure}[h]
 \centering
\includegraphics[width=14cm]{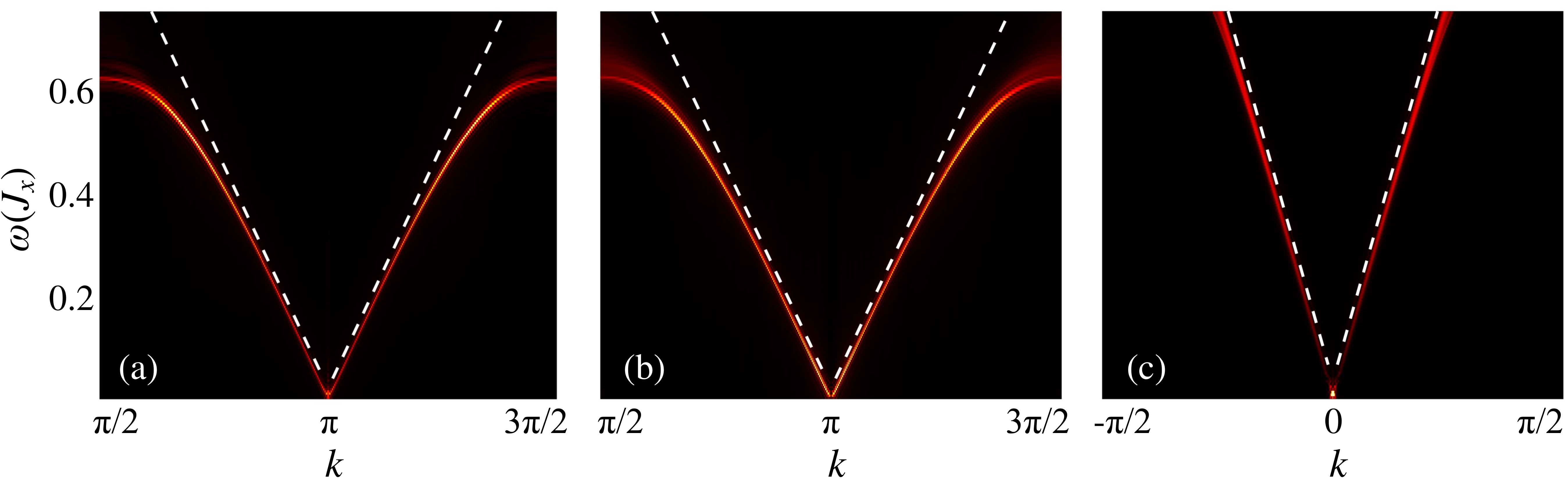}

\caption{Low-energy spectra of the dynamic current-current correlations listed
in Tab. \ref{tab:1-2}. (a), (b) and (c) correspond to correlations
of $\mathcal{J}^{c}$ at the FM-VBS transition, $\mathcal{J}^{s}$ at the FM-VBS transition
and $S^{y}$ at the $X$-$Z$ FM transition, respectively. The white lines are
fits to the linear dispersion.\label{fig:4}}
\end{figure}

As demonstrated in Fig. \ref{fig:4}(a) and (b), the the current-current
correlations in energy-momentum space at the transition points are indeed gapless with linear
low-energy dispersion. At the FM-VBS transition, the two emergent symmetries ${\rm \text{O(2)}}_{c}$and${\rm \text{O(2)}}_{s}$ can be identified from the ``charge" and ``spin" channels separately, and they
have the same Fermi velocity. Combining these results with the analysis
of anomalous exponents above, we can safely confirm that there exist
two emergent symmetries. At the DQCP of the $X$-$Z$ FM transition, $S^{y}$ is
the conserved current of the $S^{x}$-$S^{z}$ rotation as well as
of the $\Phi$-$\Gamma$ rotation. The existence of two continuous
symmetries at the $X$-$Z$ FM transition is also confirmed.

\begin{figure}[h]
 \centering
\includegraphics[width=14cm]{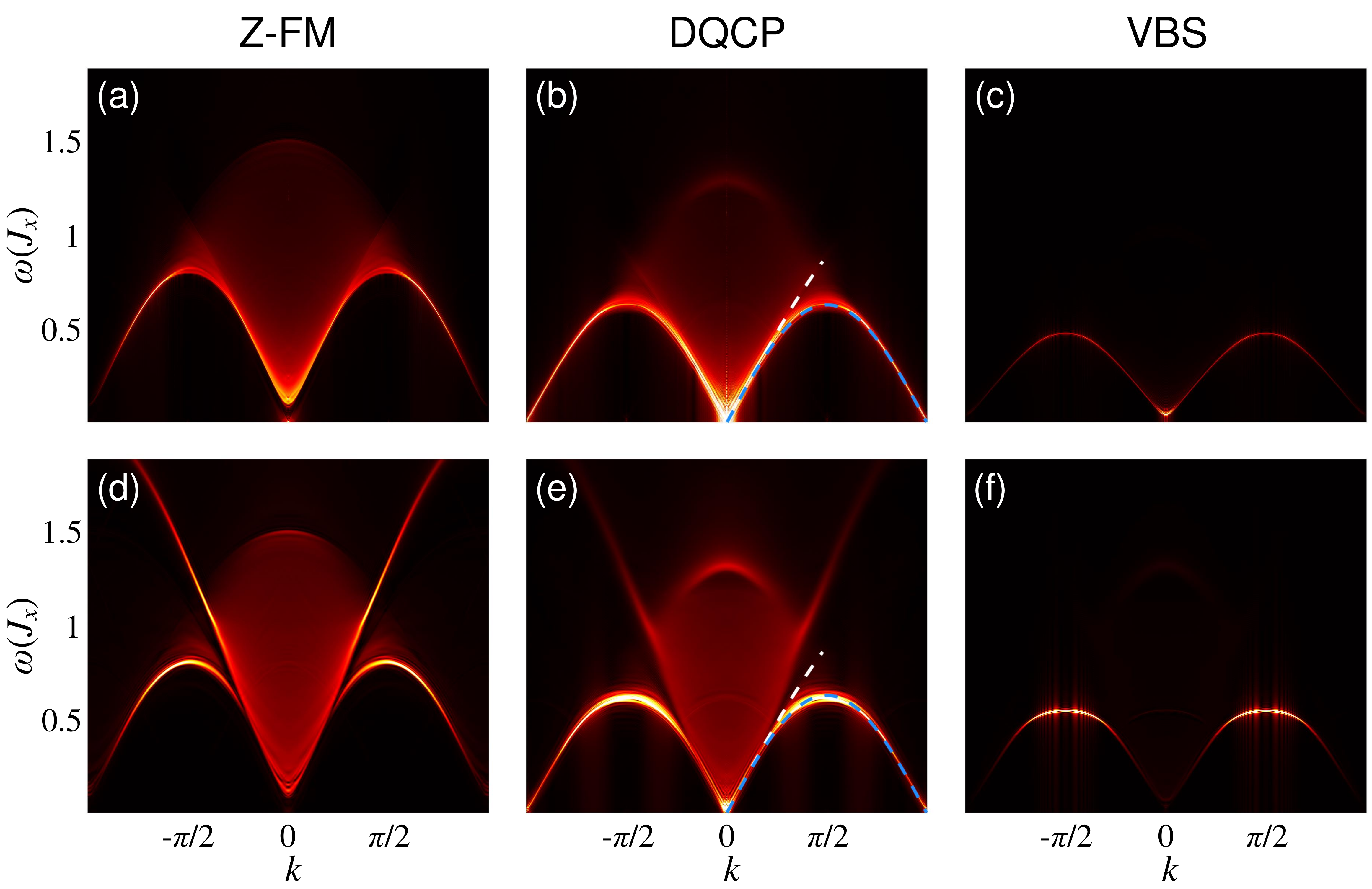}

\caption{Dynamic spin structure factors of (a-c) the $S^{z}$ channel and (d-f)
the $S^{x}$ chanel. (a) and (d) are inside the $Z$-FM phase. (b) and (e)
are at the DQCP. (c) and (f) are inside the VBS phase. The blue and
white dashed lines in (b) and (e) trace out the two bright bound edge
from (b) and are guides to eyes in (e).\label{fig:5}}
\end{figure}

The signatures of deconfined fractionalized excitations at the FM-VBS DQCP are illustrated
in Fig. \ref{fig:5} clearly. Approaching to the DQCP from either side, we observe the development of
two sets of continuous spectra (panels in the upper and lower rows of Fig. \ref{fig:5}). As shown in Fig. \ref{fig:5}(b) and (e), they
both become gapless at the DQCP.
Near the zone center, these two continua are bounded by the same edge (linear in the momentum $k$). However, when going away from the zone center by increasing $k$, they are
separated with two independent edges. Note that these two modes are associated with effective ``charge" and ``spin" channels, respectively. Therefore, this is the signature of an effective spin-charge
separation of the system. The spin-charge seperation is common in 1D Luttinger liquid. In
the long wave length limit, the interaction terms causing charge (VBS)
and magnetic ($Z$-FM) orders are all irrelavent under renormalization
group (RG) flow, which signifies free fermion excitations. However,
as the momentum $k$ increases, the effect of the scattering processes
lead to the separation of spin and charge excitations. In the case of the easy-plane DQCP in 2D proposed by Senthil et. al., the spinons (fractions of $S^{+}$ ) and merons (fractions of skyrmions) are deconfined simultaneously, in a way analogous to the spin and charge excitations discussed here.

\begin{figure}[h]
 \centering
\includegraphics[width=14cm]{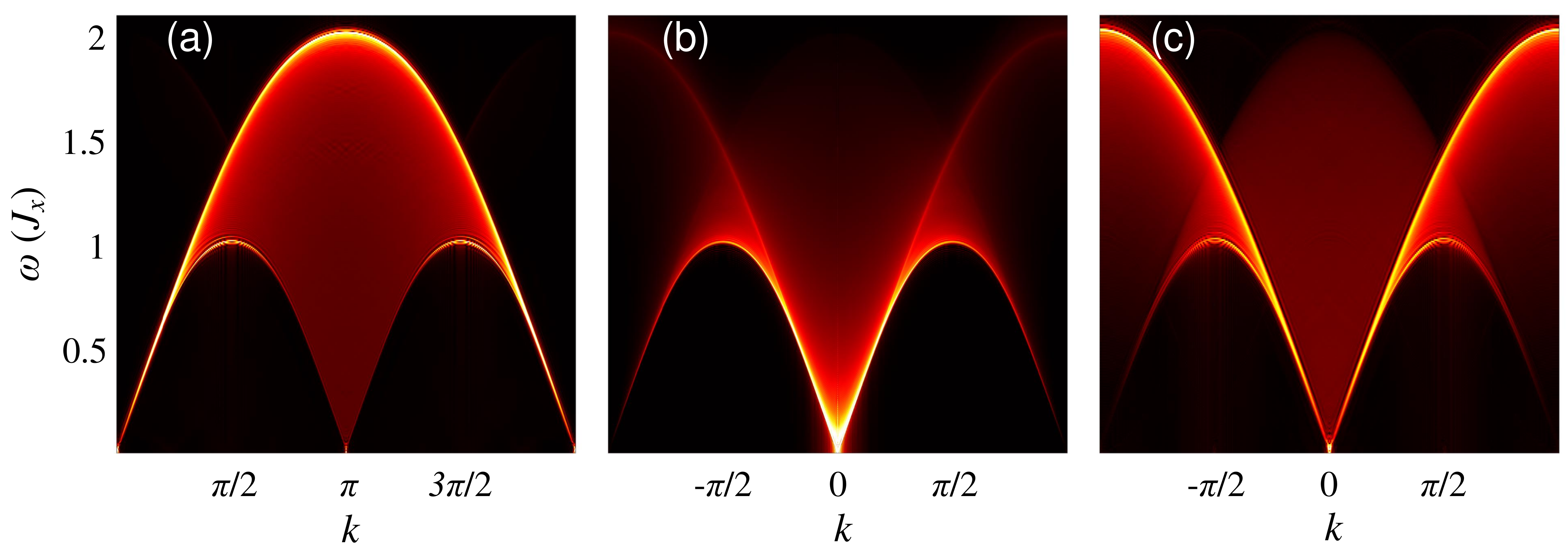}

\caption{Dynamic structure factors of (a) $S^{y}$ channel, (b) $S^{x}$ channel
, and (c) $S^{x}S^{z}$ channel at the DQCP of the $X$-$Z$ FM transition.\label{fig:6}}
\end{figure}

The excitations at the $X$-$Z$ FM DQCP are clearer. The continuum showed
in Fig. \ref{fig:6}(a) is conspicuous with clear boundaries. In the
JW fermion representation, it is nothing but the particle-hole excitations for
free fermions. The topological excitations (JW particles) are totally
deconfined in any length scale. The clear boundary is the sign of
a well defined Fermi surface. The continuum of the $S^{x}S^{z}$ channel
in Fig. \ref{fig:6}(c) can also be regarded as a measure of the topological
excitations and it is a folding of the continuum in Fig. \ref{fig:6}(a).
In the $S^{x}$ (or $S^{z}$) channel, the gapless point of the continuum
appears at $k=0$ and the boundary of the continuum is indistinct
as demonstrated in \ref{fig:6}(b), which is different from the behavior
in the $S^{y}$ channel. The excitation spectrum shown in the
$S^{x/z}$ channel suggests a Dirac cone like Fermi surface. At a first glance this is surprising because the model at this point is mapped to non-interacting JW fermions. However, we note that the JW transformation itself is nonlocal, and the $S^{x/z}$ channel just happens to probe the inherent many-body effects. This is a unique feature of 1D systems.

The anomalous exponents of $S^{x}$, $S^{y}$, and $S^{x}S^{z}$ have
been calculated in the previous section. Different $\eta$ values are also characterized in the dynamical structure factors. For
$\eta=0$, the excitation spectrum would contain a bright bound state. For $\eta=1$, which corresponds to the free JW fermions~\cite{fradkin_2013}, the spectrum should exhibit a smooth continuum just like the
one shown in Fig. \ref{fig:6}(a). And for $\eta=2$, at a fixed momentum $k$, the spectral
weight will vanish at $\omega=0$ and increase linearly with the frequency $\omega$, just like the continuum near $k=\pm\pi$
in Fig. \ref{fig:6}(c). For the correlations of $S^{x}$, $\eta=0.5$. The dynamical structure factor of this channel, as shown in Fig. \ref{fig:6}(b), exhibits a continuum with a bright edge, consisting of characters of excitations at both $\eta=0$ and $\eta=1$.

\section{Discussions and Conclusions}

As mentioned in the introduction, the realization of a DQCP in 2D
spin systems without fine-tuning is still challenging and the enormous
computation costs limit the detailed analysis of the critical properties, especially the dynamical ones. As a reasonable
instead, we numerically study the critical properties of 1D
DQCPs in detail. We show that the model in Eq.~\eqref{eq:spinmodel} contains two types of DQCPs, which are located at the VBS-FM and the $X$-$Z$ FM transitions, respectively.
They share many similar critical properties that are eligible for DQCPs, despite the different underlying mechanisms and symmetry aspects. These key differences are clearly reflected in the calculated dynamic spectra of different spin channels, which is first explored in this work. The gapless and
linear dispersions of the current-current correlations in energy-momentum
space further verify the existence of emergent symmetries. At the
VBS-FM DQCP, our calculation of dynamic spin structure factors give
robust evidence of spin-charge separation. And at the $X$-$Z$ FM DQCP,
besides the JW fermion continuum, another type of fractionalized excitation
is observed.

In the field theory description, these two different types of DQCPs can both be described by the same low-energy effective model -- the sine-Gordon model~\cite{Xi2022}, although their microscopic Hamiltonians'  symmetries are totally different. In this sense, the ground states exactly on these two lines both exhibit Luttinger liquid behavior at low energies. As expounded in Sec. 5, the low-energy excitations for these two DQCPs do not show notable difference, qualitatively. On the other hand, the spin-charge separation takes place at higher energy and larger momentum, for which the dispersions show distinct nonlinear behavior and hence beyond the field theory description. There are also other additional modes for the FM-VBS DQCPs at much higher energy. It would come as no surprise because this energy scale is far beyond the field theory capability and thus the different symmetries embodied in microscopic Hamiltonians will lead to different excitation modes. Nevertheless, the different excitation modes at higher energy may contain some interesting consequences beyond the low-energy view at present and it would inspire us to develop more analytical approaches for larger energy scale.

Our results also verify the important role of the LSM theorem played in stabilizing the 1D DQCP.
At the transition point between the $X$-FM and $Z$-FM phases, the symmetry of the Hamiltonian is enhanced to $\text{U(1)}$. Constrained by the LSM theorem, the ground state at this point should preserve the symmetry and be gapless. Although LSM theorem does not rule out a first-order transition between
the two SSB phases in general, the unique feature of 1D system excludes the possibility of a Goldstone mode. Therefore, a first-order transition with an enhanced continuous symmetry can not take place in 1D, and the system must undergoes a continuous transition, namely, via a DQCP.
As discussed, this type of DQCP can only be realized in 1D systems. In higher dimensions, emergent Goldstone mode associated with the enhanced symmetry appears and the transition is usually of first order.
On the other hand, the enhanced continuous symmetry at the DQCP between the FM and VBS phases emerges only in the long wave length. This type of DQCP is not limited in 1D.

In summary, by using the iTEBD method, we study the critical properties as well as the dynamical spectra of two types of 1D DQCP in an $S=1/2$ spin chain. The scaling of the correlations functions and the dispersion of the conserved current correlations explicitly show the emergence of enhanced continuous symmetries at the two different DQCPs. The dynamical signatures of the spin excitation spectra reveal deconfined fractionalized excitations at the DQCPs. We also find an effective spin-charge separation at the DQCP between the the FM and VBS phases, and identify two continua associated with different types of fractionalized excitations at the DQCP between the $X$-direction and $Z$-direction FM phases.
Our results uncover rich physics of the DQCP in 1D and help understand the nature of DQCP in higher dimensions.

\addcontentsline{toc}{chapter}{Appendix A: Some useful Lie algebras}
\section*{Appendix A: Some useful Lie algebras}
\begin{table}[H]
 \centering
\caption{Related SU(2) Lie algebras in this work.}
\label{tab:s1}
\begin{tabular}{c>{\centering}p{4cm}>{\centering}p{4cm}>{\centering}p{3cm}}
\toprule
\addlinespace[0.2cm]
Symmetries & \multicolumn{3}{c}{Generators}\tabularnewline\addlinespace[0.2cm]
\toprule
\addlinespace[0.2cm]
${\rm \text{SU(2)}}_{dw1}$ & $\tau_{i,j}^{1}=\text{\ensuremath{\frac{1}{2}(\sigma_{i}^{x}\sigma_{j}^{x}+\sigma_{i}^{y}\sigma_{j}^{y})}}$ & $\tau_{i,j}^{2}=\text{\ensuremath{\frac{1}{2}(\sigma_{i}^{y}\sigma_{j}^{x}-\sigma_{i}^{x}\sigma_{j}^{y})}}$ & $\tau_{i,j}^{3}=\text{\ensuremath{\sigma_{i}^{z}-\sigma_{j}^{z}}}$\tabularnewline\addlinespace[0.2cm]
\midrule
\addlinespace[0.2cm]
${\rm \text{SU(2)}}_{dw0}$ & $\nu_{i,j}^{1}=\text{\ensuremath{\frac{1}{2}(\sigma_{i}^{x}\sigma_{j}^{x}-\sigma_{i}^{y}\sigma_{j}^{y})}}$ & $\nu_{i,j}^{2}=\text{\ensuremath{\frac{1}{2}(\sigma_{i}^{y}\sigma_{j}^{x}+\sigma_{i}^{x}\sigma_{j}^{y})}}$ & $\nu_{i,j}^{3}=\text{\ensuremath{\sigma_{i}^{z}+\sigma_{j}^{z}}}$\tabularnewline\addlinespace[0.2cm]
\midrule
\addlinespace[0.2cm]
${\rm \text{SU(2)}}_{s}$ & $\sigma_{i}^{x}$ & $\sigma_{i}^{y}$ & $\sigma_{i}^{z}$\tabularnewline\addlinespace[0.2cm]
\bottomrule
\end{tabular}
\end{table}

Here we summarize some local Lie algebras in Tab. \ref{tab:s1}. They are useful in defining the order parameters and when approaching to a DQCP, global topological conservation laws arise.
The on-site spin SU(2) algebra is obvious and denoted as ${\rm \text{SU(2)}}_{s}$
in Tab. \ref{tab:s1}. Besides, there are also two sets of on-bond algebras. There are four basis states in a two-site unit cell, $\left(\left|\uparrow\uparrow\right\rangle ,\left|\downarrow\downarrow\right\rangle ,\left|\uparrow\downarrow\right\rangle ,\left|\downarrow\uparrow\right\rangle \right)$, and we can separate them
into two parts, $\left(\left|\uparrow\uparrow\right\rangle ,\left|\downarrow\downarrow\right\rangle \right)$
and $\left(\left|\uparrow\downarrow\right\rangle ,\left|\downarrow\uparrow\right\rangle \right)$,
for which the numbers of domain walls (dw) contained are 0 and 1, respectively.
The operators $\nu_{i,j}^{\alpha}$ ($\alpha=1,2,3$) defined in Tab.
\ref{tab:s1} form a closed $\text{SU(2)}$ Lie algebra in the $dw=0$ space $\left(\left|\uparrow\uparrow\right\rangle ,\left|\downarrow\downarrow\right\rangle \right)$, and is therefore denoted as ${\rm \text{SU(2)}}_{dw0}$.
While the operators $\tau_{i,j}^{\alpha}$ also form a closed $\text{SU(2)}$ algebra
in the $dw=1$ space, which is denoted as ${\rm \text{SU(2)}}_{dw1}$. All
these operators listed in Tab. \ref{tab:s1} can not mix the spaces
between $dw=0$ and $dw=1$, namely, they preserve the number of domain walls locally.

\addcontentsline{toc}{chapter}{Appendix B: Finite Entanglement Scaling}
\section*{Appendix B: Finite Entanglement Scaling}\label{Sec:AppB}
\begin{figure}
 \centering
\centering
\includegraphics[width=7.5cm]{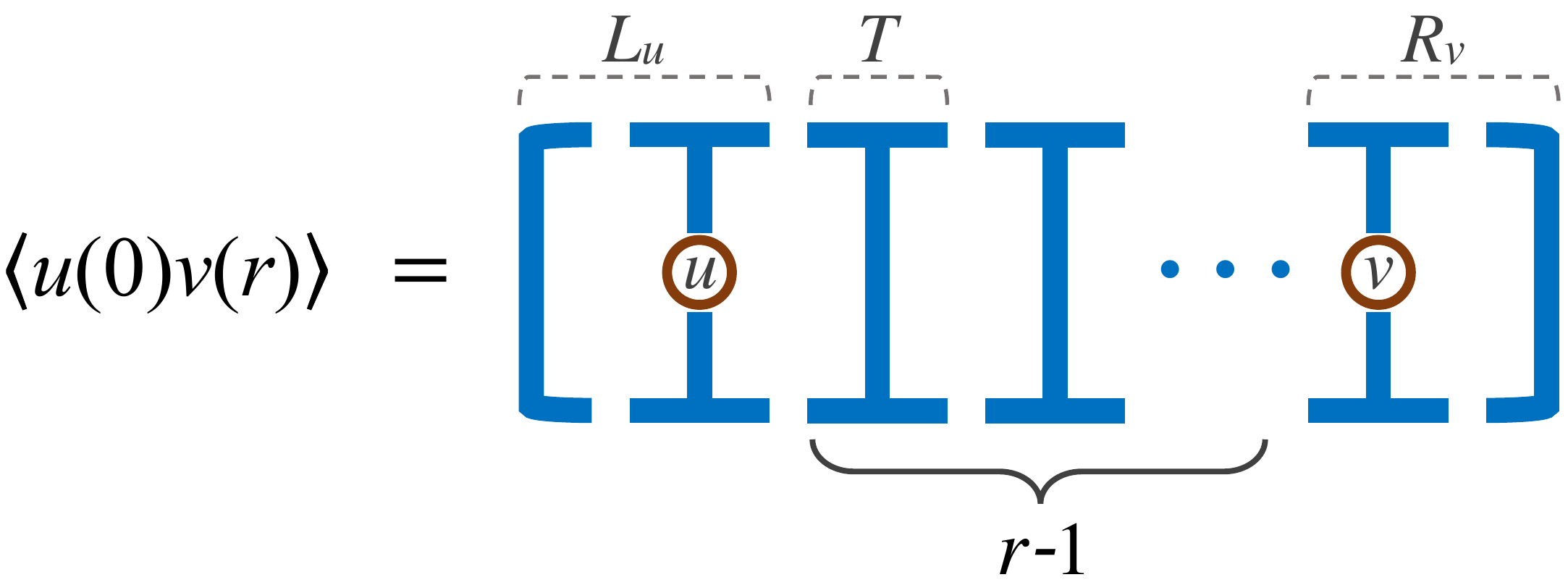}

\caption{Graph representation of two-point correlators}
\label{Fig:s1}
\end{figure}

For a matrix product state with a finite $D$, the finite gap is associated
to a finite effective correlation length $\xi(D)$. This effective correlation
length can be calculated from the correlation structures of the MPS.
The Static correlation of two operators $u$ and $v$ in the MPS language
reads~\cite{SCHOLLWOCK2011}
\begin{equation}
\begin{aligned}
\left\langle u(0)v(r)\right\rangle &=LT_{u}T^{r-1}T_{v}R \\
                                    &=L_{u}T^{r-1}R_{v}\label{eq:s1}
\end{aligned}
\end{equation}
where $T$ is the transfer operator, $L$ and $R$ are the left and
right eigenvectors of $T$ with the largest eigenvalue, respectively.
This process can be expressed in a more intelligible tensor-graph
representation as demonstrated in Fig. \ref{Fig:s1}.

Typically, the transfer operator $T$ is a symmetric matrix (corresponding
to the space-inversion symmetry) and it can be diagonalized with real
eigenvalues. Two-point correlators $\left\langle u(0)v(r)\right\rangle $
can be transformed as follows:
\begin{equation}
\begin{aligned}
\left\langle u(0)v(r)\right\rangle &=L_{u}U^{-1}\Lambda^{r-1}UR_{v} \\
                                    &=\Lambda_{1}^{r-1}\left[c_{1}+\sum_{k=2}^{rank(T)}c_{k}(\Lambda_{k}/\Lambda_{1})^{r-1}\right]\label{eq:s2}
\end{aligned}
\end{equation}
where $\Lambda_{k}$ are the eigenvalues of the transfer operator
$T$ and $\Lambda_{1}$ is the largest eigenvalue. If we transform
the transfer operator into a normalized form with $\Lambda_{1}=1$,
there remains a constant term $c_{1}$ and an exponential decay term
$\sum c_{k}\Lambda_{k}^{r-1}$ in Eq.~\eqref{eq:s2}. The constant term
guarantees the long range order. The leading decay factor $\gamma=\Lambda_{2}$ (the
second largest eigenvalue) contributes to an effective correlation length
$\xi=-1/\ln(\left|\gamma\right|)$.

If the truncation dimension $D$ is sufficiently large, the effective
correlation length $\lambda$ can be taken as the real correlation
length for a gapped state. Nevertheless, for a gapless state, this
exponential decay factor always presents, which leads to a finite-$D$ deviation from the true value at the thermodynamic limit.

The finite correlation length scaling of correlation function near
the critical point reads
\begin{equation}
G(r,\xi)=\frac{\exp(r/\xi)}{r^{\eta}}f(r/\xi)\label{eq:s3}
\end{equation}
where $\eta$ is the anomalous exponent and $f(r/\xi)$ is a homogeneous
scaling function. For the matrix product representation, the exponential
decay factor exists all along even in a gapless state. Even so, close
to the transition point, the power-law term will dominate the scaling
behavior of the correlation function for $r<\xi$.

\begin{figure}
\centering
\includegraphics[width=7.5cm]{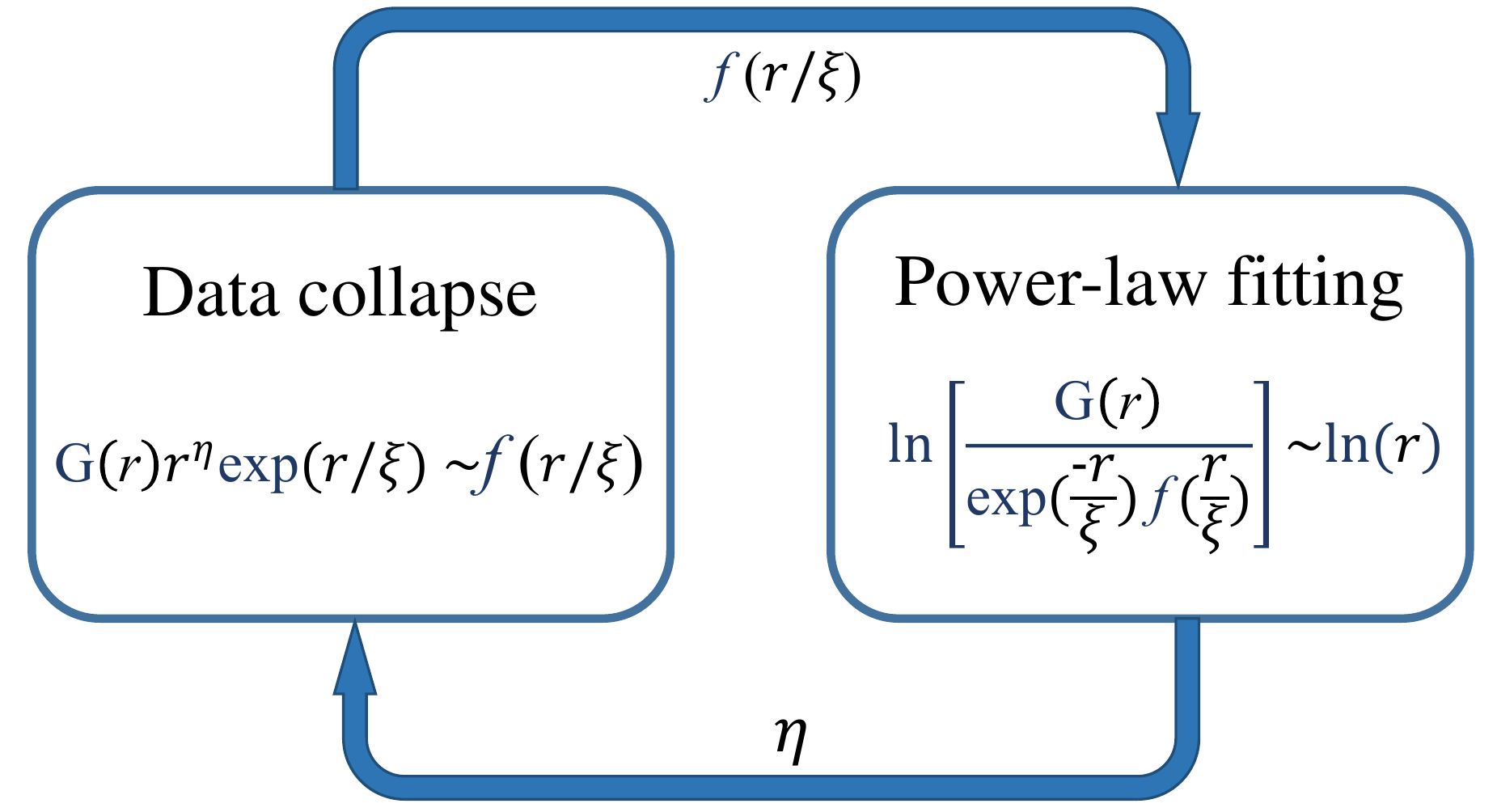}

\caption{Self-consistent process to extract the anomalous exponents.}
\label{Fig:s2}
\end{figure}

\begin{figure}
\centering
\includegraphics[width=0.7\textwidth]{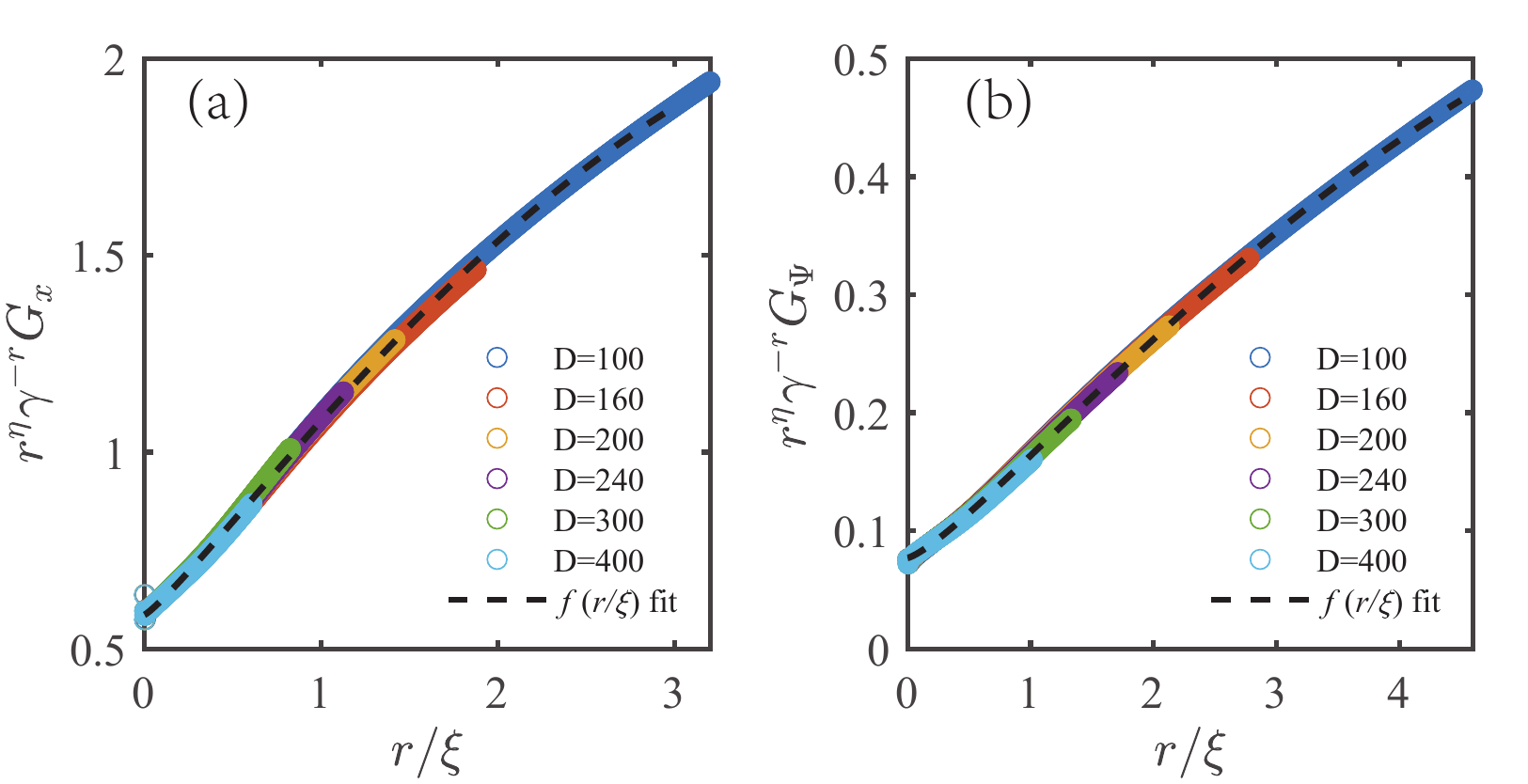}

\caption{Data collapse of the correlations functions for (a) $S^{x}$ at the $X$-$Z$ FM transition and (b) $\Psi$ at the FM-VBS transition.}
\label{Fig:s3}
\end{figure}

Data collapse has been widely adopted to extract the anomalous exponents.
However, it is often hard to judge the quality of data collapse. So it usually
only provides some rough estimate of the exponents. Alternatively, one can also
extract the anomalous exponents from a direct power-law fit to the correlation
functions. However, there are some practical issues for a direct fit
because the sub-leading contribution in the scaling function $f(r/\xi)$ would have an effect on the accuracy
of fitting for the short-range correlations and the exponential decay
factor would ruin the power-law behavior for the long-range correlations.
Here we adopt a way to extract the exponents by combining these two methods, and show that this will give refined estimates of the anomalous exponents. Specifically, direct power-law
fitting can provide an initial value of the anomalous exponent $\eta$,
and $f(r/\xi)$ fitted from data collapse can be used to promote the
power-law fitting of the correlation function. This process is then repeated until
a self consistence of $\eta$ is obtained. The sketch of our method is illustrated in Fig. \ref{Fig:s2}. The final
fitting results of the anomalous exponents are illustrated in Fig.~\ref{fig:3} of the main
text. Examples of data collapse for $G_{x}$at the $X$-$Z$ FM transition and $G_{\Psi}$ at the FM-VBS transition are shown in Fig. \ref{Fig:s3}, respectively.

\addcontentsline{toc}{chapter}{Appendix C: Calculation of dynamic structure factors}
\section*{Appendix C: Calculation of dynamic structure factors}
\begin{figure}
\centering
\includegraphics[width=0.7\textwidth]{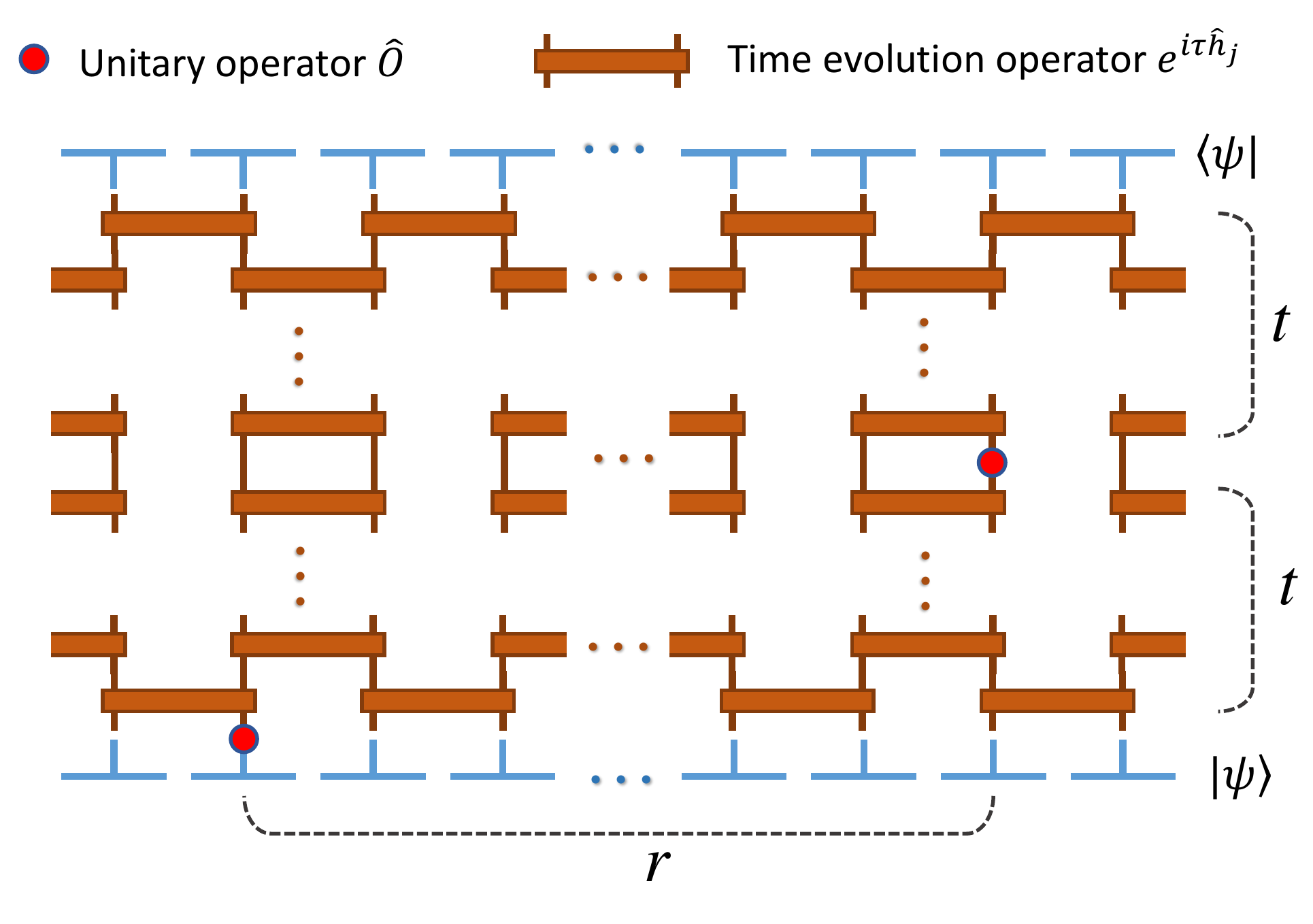}

\caption{Sketch of calculating the space-time correlation $\left\langle \hat{O}(0,0)\hat{O}(r,t)\right\rangle $
in the matrix product representation.}
\label{Fig:s4}
\end{figure}

In this section, we briefly describe the process of calculating the dynamical structure factors in the matrix product representation. We first calculate the zero-temperature space-time correlation $\left\langle \hat{O}(0,0)\hat{O}(r,t)\right\rangle $
by using the iTEBD method. The calculation of the space-time correlation of unitary operators is a standard operation
of iTEBD. The space-time correlation can be written as
\begin{equation}
\begin{aligned}
\left\langle \hat{O}(0,0)\hat{O}(r,t)\right\rangle &= \left\langle \psi_{G}\right|\hat{O}(0)e^{-it\hat{H}}\hat{O}(r)e^{it\hat{H}}\left|\psi_{G}\right\rangle \\
&=\left\langle \psi_{L}\right|e^{-it\hat{H}}\hat{O}(r)e^{it\hat{H}}\left|\psi_{G}\right\rangle.\label{eq:s4}
\end{aligned}
\end{equation}
If the ground state $\left\langle \psi_{G}\right|$ is in the canonical
form, the state $\left\langle \psi_{L}\right|=\left\langle \psi_{G}\right|\hat{O}(0)$
will also be in the canonical form, for $\hat{O}$ is a unitary operator.
Then standard operations of the real-time evolution can be applied as
shown in Fig. \ref{Fig:s4}.
After that, dynamical structure factor of $\hat{O}$ can be obtained
by Fourier transformation of $\left\langle \hat{O}(0,0)\hat{O}(r,t)\right\rangle $.

In the practical calculation, a fourth-order Suzuki-Trotter decomposition method is applied to eliminate the time-step errors as possible. The values of the time step $\tau$ and number of steps $N$ are experience-based decisions. A larger $\tau$ will result in a larger Trotter error and a smaller range of energy. There are also some practical problems to improve the energy resolution by taking a large number of steps. The truncation errors will increase with larger $N$, and the memory cost will grow quadratically with $N$. To keep balance between the computing resource and acceptable errors, in the calculation of the space-time correlations, we take the truncation dimension $D=48$, the time step $\tau=0.6J_x^{-1}$(more accurate than $\tau=0.04J_x^{-1}$ by second-order decomposition), and the number of steps $N=1200$.

To trace out the low-energy linear dispersion in $k$-$\omega$ space, we find out the $\omega$ values corresponding to the maximum of DSF for each $k$ and then fit the $k$-$\omega$ data nearby the zoom center linearly. The linear guide lines in the main text are our linear-fitting results in this way and the boundary guide lines are direct plots of $k$-$\omega$ data.

\addcontentsline{toc}{chapter}{Acknowledgment}
\section*{Acknowledgment}
We thank Z.~Y. Xie, Z.-X. Liu, W.~Q. Yu, Y. Wang and C.~L. Liu for useful discussions. Financial supports are given in the footnote on the first page.


\begin{thebibliography}{99}
\bibitem{entanglefootnote}
For a system with open boundary conditions, the scaling of the entanglement
  entropy reads $s\sim\frac{c}{6}log(\xi)$, where $c$ is the central charge.
\end{thebibliography}
\end{document}